\documentclass[onecolumn,amsmath,amssymb,nofootinbib,12pt]{article}
\usepackage{jheppub}
\usepackage{ifpdf}
\usepackage{graphicx}
\usepackage{amsfonts}
\usepackage{amsmath}
\usepackage{amssymb}
\usepackage{epsfig}
\usepackage{pdfpages}
\usepackage{graphicx,epstopdf}

\usepackage{caption}
\usepackage[position=b]{subcaption}

\usepackage[makeroom]{cancel}
\usepackage{hyperref}
\hypersetup{pdftex,colorlinks=true,allcolors=blue}
\usepackage{array}

\def\pa{\partial}

\newcommand{\br}{\biggr}
\newcommand{\bl}{\biggl}

\begin{document}


\title{Periapsis shift and deflection of light by hairy black holes}

\author[a,b]{David Choque,}
\author[a]{Willson D. Llerena,}
\author[a]{Jhiwsell D. Vargas}
\affiliation[a]{Universidad Nacional de San Antonio Abad del Cusco, Av. La Cultura 733, Cusco, Per\'u.}
\affiliation[b]{Universidad T\'{e}cnica Federico Santa Mar\'{\i}a, Av. Espa\~{n}a 1680, Valpara\'{\i}so, Chile.}

\emailAdd{brst1010123@gmail.com}
\emailAdd{120252@unsaac.edu.pe}
\emailAdd{j.dennis.29.2@gmail.com}

\date{\today}

\abstract{We investigate the orbit equations and the eikonal equation for light respectively, under influence of the hairy black holes (asymptotically flat) in fourth dimensions. We consider two hairy black hole solutions with non-trivial potentials, and one of these solutions has Schwarzschild case as a smooth limit. Following to Landau and Lifshitz we use the Hamilton-Jacobi method, and we show hairy corrections for periapsis shift, where the effect of the hair is to increase it. In the same way, using the eikonal equation we show the deflection of the light and the relevant scalar hair corrections. Interestingly we find that the hair screening the gravitational field, decreasing the angle of deflection as the hair increases.}    

\maketitle

\section{Introduction}
%
The Hamilton-Jacobi formalism  
was applied in different gravity situations,
for example in the context of AdS/CFT \cite{Maldacena:1997re}
it was used to find
counterterms \cite{Martelli:2002sp,Batrachenko:2004fd}.
On the other hand, the Hamilton-Jacobi equations of canonical gravity are the main tool in the derivation of renormalization group equations \cite{deBoer:2000cz,Heemskerk:2010hk}.

The scalar field has a fundamental
structure in string theory, and the
expectation value controls the 
string coupling constant $g_{s}=\langle e^{\phi}\rangle$. In high energy
physics the scalar field is the 
Higgs particle \cite{Aad:2012tfa}, and in Cosmology is the $inflaton$ that
describes cosmological perturbations
observed in the experiments COBE, WMAP and Planck \cite{Spergel:2006hy,Planck:2013jfk}. 
In the context of
AdS/CFT the scalar field is a source 
of the boundary operators\footnote{This depends on Dirichlet, Newmann or mix boundary conditions of the scalar field \cite{Anabalon:2015xvl,Henneaux:2006hk}.}.  
Interestingly the hairy solutions studied here can be obtained from his asymptotically AdS solution. By suitably adjusting the cosmological constant or the scalar field potential so that the effective cosmological vanishes, one can obtain asymptotically flat black holes \cite{Anabalon:2013qua}. We considered the boundary
condition of a scalar field such that $\phi=0$ (at spatial infinity)\footnote{ 
	In \cite{Gibbons:1996af,Astefanesei:2006sy} considered another interesting boundary condition in which the moduli have a non-trivial radial dependence, where the properties of these black holes depend on the values $\phi_{\infty}$ of moduli at spatial infinity.}.   
In \cite{Anabalon:2013qua,Anabalon:2012ih,Gibbons:1987ps} was constructed various exact and regular solutions of hairy black holes (charged) asymptotically flat, and AdS \cite{Anabalon:2012ta,Anabalon:2012ih,Anabalon:2012dw}. Is important understanding that the hair (scalar field) lives behind horizon and boundary, and 
does not exist a conserved quantity associated with the scalar field. In
\cite{Anabalon:2013qua,Nunez:1996xv} argument that the back-reaction
ensure that the scalar field
can over in a strong gravitational field without collapsing completely. In the present paper, we considered 
Einstein-dilaton theories minimally coupled and with non-trivial potentials,
in \cite{Nunez:1996xv,Anabalon:2012ih,Anabalon:2012dw} constructed hairy black holes
in dilaton-charged gravity theories without 
scalar potential, in these cases 
the coupling of the dilaton  
with gauge fields ensures the existence of 
an effective potential. In the extremal case the near horizon data depend completely by the electric and magnetic charges and so the attractor mechanism \cite{Ferrara:1995ih,Strominger:1996kf,Ferrara:1996dd}  works like the no-hair theorem.\\

To describe the geodesics (and null geodesic) we need to consider
two actions, one for the gravitational
field which describes the geometric of
space-time for a given distribution of matter $T_{\mu\nu}$. The another action 
gives us information about how the particles (of mass $m$) moves in that space-time and is given by
\begin{equation}
I=-mc\int_{a}^{b}ds
\end{equation} 
where $ds$ is the infinitesimal 
line universe $ds^{2}=g_{\mu\nu}dx^{\mu}dx^{\nu}$. Taking the variation we get the geodesic equation
\begin{equation}
\frac{d^{2}x^{\mu}}{ds^{2}}+\Gamma_{\alpha\beta}^{\mu}\frac{dx^{\alpha}}{ds}\frac{dx^{\beta}}{ds}=0
\end{equation}
but to describe the null geodesics
($ds=0$) the before equation 
is not complete\footnote{That problem can solve easily, but we consider another option.}.
The Hamilton-Jacobi is a powerful
method to obtain the orbit equations. Landau-Lifshitz 
showed that the Hamilton-Jacobi equation for a particle in any space-time is $(\mathcal{M},g)$
\begin{equation}
g^{\mu\nu}\frac{\pa I}{\pa x^{\mu}}\frac{\pa I}{\pa x^{\mu}}-m^{2}=0
\end{equation}
In literature there is an extensive bibliography about the study of the orbits of several black holes in different theories \cite{Frye:2013xia,Cruz:2004ts,Magnan:2007uw,Cruz:2011yr,Wells:2011st,Hackmann:2008zz,Olivares:2013zta}.
Recently was study Light propagation in a plasma on Kerr space-time using 
the Hamilton-Jacobi equation \cite{Perlick:2017fio}, but 
in fourth dimensions is complicated
find a rotated hairy black holes, although there are currently several solutions in three dimensions \cite{Correa:2012rc,Natsuume:1999at}.\\
\\
For light trajectory (null geodesic)
we consider the null module 
condition for four-vector wave
$k_{\mu}k^{\mu}=0$. So, replacing 
$k_{\mu}=\pa\psi/\pa x^{\mu}$ 
we find eikonal equation in 
a gravitational field
\begin{equation}
g^{\mu\nu}\frac{\pa \psi}{\pa x^{\mu}}\frac{\pa \psi}{\pa x^{\mu}}=0
\label{iconal}
\end{equation}
The present paper is structured as follows: In section 2, we consider a small review of Newtonian results and its corrections given by general relativity. In section 3, we present the two hairy solutions and its properties, in the section 4 we use the holographic stress tensor method to determine the mass of hairy black holes. Finally, in section 5, following to intuitive procedure of section 2, we calculate the periapsis shift and deflection of light for both hairy black holes, and in the final part, we present the conclusions.     
%
\section{Hamilton-Jacobi method in Schwarzschild}
The Kepler problem in celestial mechanics consist in found and solve the orbit equations
of stellar systems (two body system). Classically is
solved (exactly) considering the Newtonian  potential
\begin{equation}
U_{N}(r)=-\frac{G_{N}mm^{'}}{r}
\end{equation}
And we can be solved alternatively
by Hamilton-Jacobi method.
Considering the Lagrangian  expression for
a particle in the central field force (at the plane $\theta=\pi/2$)
\begin{equation}
L=\frac{m}{2}(\dot{r}^{2}+r^{2}\dot{\varphi}^{2})-U_{N
}(r), \qquad M=\frac{\pa L}{\pa\dot{\varphi}}=mr^{2}\dot{\varphi}
\end{equation}
The Hamilton-Jacobi equation is
\begin{equation}
\frac{1}{2m}\bl{(}\frac{\pa I}{\pa r}\br{)}^{2}+\frac{1}{2mr^{2}}\bl{(}\frac{\pa I}{\pa \varphi}\br{)}^{2}+U_{N}(r)=\mathcal{E}^{'}
\end{equation}
where the ansatz is $I(r)=-\mathcal{E}^{'}t+M\varphi+I_{r}^{(0)}$ and its solution is
\begin{equation}
I(r)=-\mathcal{E}^{'}t+M\varphi+\int{\sqrt{2m(\mathcal{E}^{'}-U_{N})-\frac{M^{2}}{r^{2}}}dr}
\label{Ir}
\end{equation}
The trajectory is given by equation $\frac{\pa I}{\pa M}=cte$
\begin{equation}
\varphi(r)=\int\frac{Mdr}{r^{2}\sqrt{2m(\mathcal{E}^{'}-U_{N})-\frac{M^{2}}{r^{2}}}}+\varphi_{0} \Rightarrow \frac{p}{r}=1+e\cos{(\varphi-\varphi_{0})}
\label{phi0}
\end{equation}
Where $p=M^{2}/m^{2}m^{'}G_{N}$ and the eccentricity is $e=\sqrt{1+2\mathcal{E}^{'}p/G_{N}mm^{'}}$. Is easy to show that for elliptical orbit 
\footnote{In this case we have $\mathcal{E}^{'}<0$ and  $r_{min}=p/(1+e)=a(1-e)$, $r_{max}=p/(1-e)=a(1+e)$.} $e<1$
\begin{equation}
\Delta\varphi^{(0)}=-\frac{\pa \Delta I_{r}^{(0)}}{\pa M}=2\int_{r_{min}}^{r_{max}}\frac{Mdr}{r^{2}\sqrt{2m(\mathcal{E}^{'}-U_{N})-\frac{M^{2}}{r^{2}}}}=2\pi
\label{phichan}
\end{equation}
This is the angle that vector position
tour when $r$ changed from $r_{max}$ to $r_{min}$ and return to $r_{max}$\footnote{Another interesting result is $\mathcal{E}^{'}=-G_{N}mm^{'}/2a$, where $m^{'}$ is the mass of the star, black hole or another big gravitational source. And $a$ is the length of the semi-major axis of the elliptical orbit.}. 
The astronomical observations showed that perihelion of Mercury has deviations $\Delta\varphi^{(0)}\rightarrow \Delta\varphi^{(0)}+\delta\varphi$~\footnote{The closest point of the celestial object that orbits (closed orbit) another is known as $periapsis$, for a particular case of the solar system we use $perihelion$.}. Exist
different forms to deal with this problem, classically we can consider 
perturbations to Newtonian potential 
$U_{N}\rightarrow U_{N}+\delta U(r)$ \cite{Landau,LandauL,Wells:2011st}.
Another option is to consider the corrections of general relativity. The usual method consist in to solve the geodesic equation \cite{Magnan:2007uw}, but here we focus on the Hamilton-Jacobi method \cite{Vasudevan:2005js}.
%
\subsection{General relativistic corrections}
\label{Gr}
In this section, we describe step by step the procedure proposed by Landau and Lifshitz \cite{Landau}. This gives us a correct intuition when we face the hairy case. We will consider the very know Schwarzschild asymptotically flat solution, where the action is
\begin{equation}
S[g_{\mu\nu}]=\frac{1}{2\kappa}\int_{\mathcal{M}}{d^{4}x\sqrt{-g}R}+\frac{1}{\kappa}%
\int_{\partial\mathcal{M}}{d^{3}xK\sqrt{-h}} \label{actionShw}%
\end{equation}
Here $\kappa=8\pi G_{N}$, and
the last term is the Gibbons-Hawking boundary term. Where $h_{ab}$
is the boundary metric and $K$ is the trace of the extrinsic
curvature. The solution is
\begin{equation}
ds^{2}=-c^2 N(r)dt^{2}+\frac{dr^{2}}{N(r)}+r^{2}(d\theta^{2}+\sin^{2}{\theta}d\varphi^{2})
\end{equation}
where, $N(r)=1-r_{g}/r$ and $r_{g}=2G_{N}m^{'}/c^{2}$. Here $m^{'}$ is the gravitational mass.
Considering geodesics in the plane $\theta=\pi/2$. The Hamilton-Jacobi
equation of the trajectory is
\begin{equation}
\frac{1}{c^{2}N(r)}\bl{(}\frac{\pa I}{\pa t}\br{)}^{2}
-N(r)\bl{(}\frac{\pa I}{\pa r}\br{)}^{2}-\frac{1}{r^{2}}
\bl{(}\frac{\pa I}{\pa \varphi}\br{)}^{2}-m^{2}c^{2}=0
\label{HJ1}
\end{equation}
The ansatz is $I(t,r,\varphi)=-\mathcal{E}_{0}t+M\varphi+I_{r}(r)$,
where the energy and mass of particle is $\mathcal{E}_{0}$, $m$, and its
angular momentum is $M$. Replacing the expression $I(t,r,\varphi)$ in (\ref{HJ1})
we can solve $I_{r}$
\begin{equation}
I_{r}(r)=\int{\bl{[}\frac{r^{2}(\mathcal{E}_{0}^{2}-m^{2}c^{4})+m^{2}c^{4}rr_{g}}{c^{2}(r-r_{g})^2}-\frac{M^{2}}{r(r-r_{g})}\br{]}}^{1/2}dr
\label{IGR}
\end{equation} 
The trajectory equation can be 
determined by $\varphi(r)=-\frac{\pa I_{r}}{\pa M}+cte$,
and is easy to show
\begin{equation}
\varphi(r)=\int \bl{[}r^{2}\sqrt{\frac{\mathcal{E}_{0}^{2}}{c^{2}}-\bl{(}m^{2}c^{2}+\frac{M^{2}}{r^{2}}\br{)}\bl{(}1-\frac{r_{g}}{r}\br{)}}\br{]}^{-1}Mdr+cte
\label{phi1}
\end{equation}
The classical limit given in (\ref{phi0}) can be obtained considering $c\rightarrow\infty$, $r_{g}=0$ and  $\mathcal{E}_{0}=mc^{2}+\mathcal{E}^{'}$. Here $\mathcal{E}^{'}$ is the non-relativistic energy. To obtain the relevant corrections of general relativity we consider small velocities of the planets respect to light\footnote{For example, the star S2 that orbit to Sagittarius $A^{*}$ (super massive black hole) in the periapsis the velocity is five percent of the light velocity \cite{Schodel:2002vg}.}, this means $r_{g}/r<<1$. The angular change of closed orbits can be calculated in the similar form to (\ref{phichan})
\begin{equation}
\Delta\varphi(r)=-\frac{\pa\Delta I_{r}}{\pa M}
\label{varphii}
\end{equation}   
For that purpose, we need expand (\ref{IGR}) and check the Newtonian term and its relativistic correction. Comparing $M^{2}/r^{2}$ part of (\ref{Ir}) and (\ref{IGR}) is easy concluded that we need to consider the change variable
\begin{equation}
r(r-r_{g})=r^{'2} \Rightarrow r-\frac{r_{g}}{2}\approx r^{'}
\label{newcoor}
\end{equation}
and introducing the non-relativistic energy $\mathcal{E}^{'}$. Expanding around $r_{g}/r<<1$ keeping fix $M^{2}/r^{2}$ (regardless of the apostrophe) we have
\begin{equation}
I_{r}=\int\sqrt{2m(E-U)-\frac{M^{2}}{r^{2}}+\frac{3m^{2}c^{2}r_{g}^{2}}{2r^{2}}+O\bl{(}\frac{r_{g}^{3}}{r^{3}}\br{)}}dr
\end{equation}
where
\begin{equation}
E=\mathcal{E}^{'}+\frac{\mathcal{E}^{'2}}{c^{2}}, \qquad U=U_{N}\bl{(}1+\frac{4\mathcal{E}^{'}}{mc^{2}}\br{)}, \qquad U_{N}=-\frac{mm^{'}G_{N}}{r}
\label{EUN}
\end{equation}
We can see the fundamental correction to $M^{2}/r^{2}$ is given by the term $r_{g}^{2}/r^{2}$, and this term can describe the periapsis shift of the orbit. We consider $E\approx\mathcal{E}^{'}$, $U\approx U_{N}$ and only expanding around $
r_{g}/r<<1$ keeping fix $M^{2}/r^{2}$
\begin{equation}
I_{r}\approx\int\sqrt{2m(\mathcal{E}^{'}-U_{N})-\frac{M^{2}}{r^{2}}}~dr+\frac{3m^{2}c^{2}r_{g}^{2}}{4}\int\frac{1}{r^{2}}\bl{[}\sqrt{2m(\mathcal{E}^{'}-U_{N})-\frac{M^{2}}{r^{2}}}\br{]}^{-1}dr+\ldots
\end{equation}
Considering the limits of integration $(r_{min}, r_{max})$, we have $I_{r}\rightarrow \Delta I_{r}$
\begin{equation}
\Delta I^{(0)}_{r}=\int_{r_{min}}^{r_{max}}\sqrt{2m(\mathcal{E}^{'}-U_{N})-\frac{M^{2}}{r^{2}}}~dr
\label{deltaI0}
\end{equation}
So~\footnote{You can see that (\ref{deltaI0}) is the same expression given in (\ref{Ir}).}
\begin{equation}
\Delta I_{r}\approx\Delta I_{r}^{(0)}-\frac{3m^{2}c^{2}r_{g}^{2}}{4M}~\frac{\pa\Delta I_{r}^{(0)}}{\pa M}
\end{equation} 
then, using (\ref{varphii}) is easy to show the correction
given for general relativity\footnote{For perihelion shift of Mercury this give $43^{''}$ and astronomical observations give $43,1^{''}\pm 0,4^{''}$.}
\begin{equation}
\Delta\varphi\approx2\pi+\frac{6\pi G_{N}^{2}m^{2}m^{'2}}{c^{2}M^{2}}
\label{Schwaphi}
\end{equation}
To determine the trajectory of light ray (null geodesics)
we need the eikonal equation (\ref{iconal}).
This equation is similar to (\ref{HJ1}) with $m^{2}=0$ and $I\rightarrow\psi$, and in this 
case, instead of energy $\mathcal{E}=-\pa I/\pa t$ of the particle, we consider the light
frequency $\omega_{0}=-\pa\psi/\pa t$, then
\begin{equation}
\psi(r)=-\omega_{0}t+M\varphi+\psi_{r}(r)
\end{equation}  
From eikonal equation we have
\begin{equation}
\psi_{r}(r)=\frac{\omega_{0}}{c}\int
\bl{[}\frac{r^{2}}{(r-r_{g})^{2}}-\frac{\varrho^{2}}{r(r-r_{g})}\br{]}^{1/2}dr, \qquad \varrho=\frac{Mc}{\omega_{0}}
\end{equation}
where $\varrho$ is the impact parameter. One more time we need consider the before transformations given in (\ref{newcoor})
\begin{equation}
\psi_{r}(r)\approx\frac{\omega_{0}}{c}\int\bl{(}1+\frac{2r_{g}}{r}-\frac{\varrho^{2}}{r^{2}}\br{)}^{1/2}dr
\label{psiSchw}
\end{equation} 
the equation for the trajectory of a light ray
is $\varphi(r)=-\frac{\pa\psi_{r}}{\pa M}+cte$,
then
\begin{equation}
\varphi(r)=\int\frac{1}{r^{2}}\bl{[}\sqrt{\frac{1}{\varrho^{2}}-\frac{1}{r^{2}}\bl{(}1-\frac{r_{g}}{r}\br{)}} \br{]}^{-1}dr+cte
\label{lightrayec}
\end{equation}
the gravitational correction is given by $r_{g}$, if we take $r_{g}=0$ the integral is $r=\varrho/\cos{\varphi}$. Namely a line that passing a distance $\varrho$ from the origin, integrating
(\ref{psiSchw}) when $r_{g}=0$
\begin{equation}
 \Delta\psi_{r}^{(0)}=\frac{\omega_{0}}{c}\int_{0}^{\pi}\sqrt{1-\frac{\varrho^{2}}{r^{2}}}dr=-M\pi
\end{equation}
The angle formed by the asymptotes of ray light that comes from a very great distance $(r\rightarrow\infty, \varphi=-\pi/2)$, approaches the nearest point $(r=\varrho, \varphi=0)$ and moves away a great distance $(r\rightarrow\infty, \varphi=\pi/2)$, is $\Delta\varphi^{(0)}=-\pa\Delta\psi_{r}^{(0)}/\pa M=\pi$. The relativistic corrections are given by
\begin{equation}
\Delta\varphi=-\frac{\pa\Delta\psi_{r}}{\pa M}
\label{light}
\end{equation}
and for that purpose we need expand
(\ref{psiSchw}) around $r_{g}/r<<1$ (keeping constant $\varrho^{2}/r^{2}$)
\begin{equation}
\psi_{r}(r)\approx \psi^{(0)}_{r}+\frac{r_{g}\omega_{0}}{c}\ln{\varrho}+\frac{r_{g}\omega_{0}}{c}\cosh^{-1}\bl{(}\frac{r}{\varrho}\br{)}
\end{equation}
The term $\psi^{(0)}_{r}$ corresponds to the classical ray rectilinear $\psi_{r}(r_{g}=0)=\psi^{(0)}_{r}$ in (\ref{psiSchw}). The total changing of $\psi_{r}$ during propagation of the light, from one large distance $R$ at the point $r=\varrho$ close to the center and back again to the distance $R$ is
\begin{equation}
\Delta\psi_{r}=\Delta\psi^{(0)}_{r}+
\frac{2r_{g}\omega_{0}}{c}\cosh^{-1}\bl{(}\frac{r}{\varrho}\br{)}
\end{equation}
replacing in (\ref{light})
\begin{equation}
\Delta\varphi=-\frac{\pa\Delta\psi^{(0)}_{r}}{\pa M}+\frac{2r_{r}R}{\varrho\sqrt{R^{2}-\varrho^{2}}}=\Delta\varphi^{(0)}+\frac{2r_{r}R}{\varrho\sqrt{R^{2}-\varrho^{2}}}
\end{equation}
taking the limit $R\rightarrow\infty$
and remember that $\Delta\varphi^{(0)}=\pi$
for the rectilinear ray, we obtain
\begin{equation}
\Delta\varphi=\pi+\frac{2r_{g}}{\varrho}
\label{Schwalighdeflex}
\end{equation}
\newpage
\section{Hairy black hole solutions}
We will consider two hairy solutions one of which is 
the smooth limit of the first 
when the hairy parameter $\nu\rightarrow\infty$.  
We are interested in asymptotically flat hairy black hole solutions with a
spherical horizon \cite{Anabalon:2013sra, Anabalon:2013eaa}. The action is
\begin{equation}
S[g_{\mu\nu},\phi]=\frac{1}{2\kappa}\int_{\mathcal{M}}{d^{4}x\sqrt{-g}\biggl{[}R%
	-\frac{(\partial\phi)^{2}}{2}-V(\phi)\biggr{]}}+\frac{1}{\kappa}%
\int_{\partial\mathcal{M}}{d^{3}xK\sqrt{-h}}+S^{ct} \label{action}%
\end{equation}
where $V(\phi)$ is the scalar potential, $\kappa=8\pi G_{N}$ and $S^{ct}$ is the boundary counterterm which we use for constructed the renormalized quasi-local stress tensor $\tau_{ab}$.
The
equations of motion for the dilaton and metric are
\[
\frac{1}{\sqrt{-g}}\partial_{\mu}\left(  \sqrt{-g}g^{\mu\nu}\partial_{\nu}%
\phi\right)  -\frac{\partial V}{\partial\phi}=0
\]
\[
E_{\mu\nu}=R_{\mu\nu}-\frac{1}{2}g_{\mu\nu}R-\frac{1}{2}T_{\mu\nu}^{\phi}
\]
where the stress tensor of the scalar field is
\[
T_{\mu\nu}^{\phi}=\partial_{\mu}\phi\partial_{\nu}\phi-g_{\mu\nu}\left[
\frac{1}{2}\left(  \partial\phi\right)  ^{2}+V(\phi)\right]
\]
When we consider convex or positive semi-definite potential the no-hair theorems ensure that does not exist regular black holes solutions \cite{Israel:1967wq,Bekenstein:1995un,Sudarsky:1995zg}. In \cite{Martinez:2006an,Acena:2013jya,Anabalon:2013sra,Anabalon:2013qua,Acena:2012mr,Anabalon:2012dw} they were able to relax some of those conditions and find a large (exact) family of hairy black holes. 
In the present article, we will work 
with following potentials, where
$l_{\nu}^{-1}=\sqrt{(\nu^{2}-1)/2\kappa}$ and $j^{-1}=1/\sqrt{2\kappa}$
\footnote{In \cite{Anabalon:2013eaa,DallAgata:2012mfj,deWit:2013ija} 
	was showed that
	these scalar potentials (\ref{pot1}), (\ref{pot2}) for asymptotically-AdS space-times, fixing some particular values of the parameters it becomes the one of a truncation of $\omega$-deformed gauged $\mathcal{N}=8$ supergravity.} 
\begin{align}
V(\phi)  &  =\frac{2\alpha}{\nu^{2}}\biggl{[}\frac{\nu-1}{\nu+2}\sinh{\phi l_{\nu
	}(\nu+1)}-\frac{\nu+1}{\nu-2}\sinh{\phi l_{\nu}(\nu-1)}+4\frac{\nu^{2}-1}%
{\nu^{2}-4}\sinh{\phi l_{\nu}}\biggr{]}
\label{pot1}
\end{align}
\begin{equation}
V(\phi)=\frac{\alpha j\phi}{\kappa}[2+\cosh
(j\phi)]-\frac{3\alpha}{\kappa}\sinh(j\phi)
\label{pot2}
\end{equation}
Along the present paper, we centered in the negative branch: $\phi\in (-\infty,0]$ for which $\alpha>0$ 
\begin{figure}[h]
	\centering
	\includegraphics[scale=0.358]{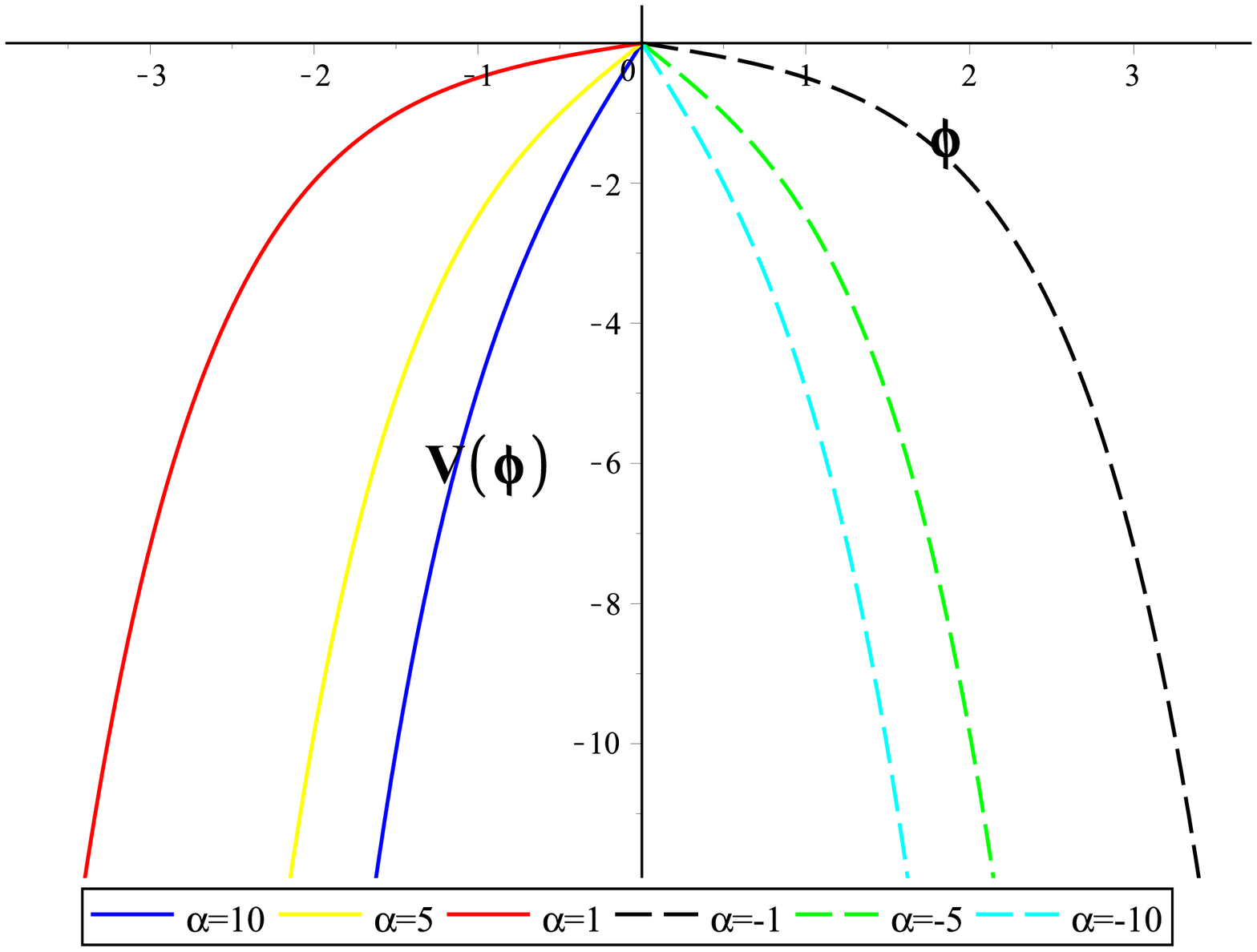}
	\includegraphics[scale=0.31]{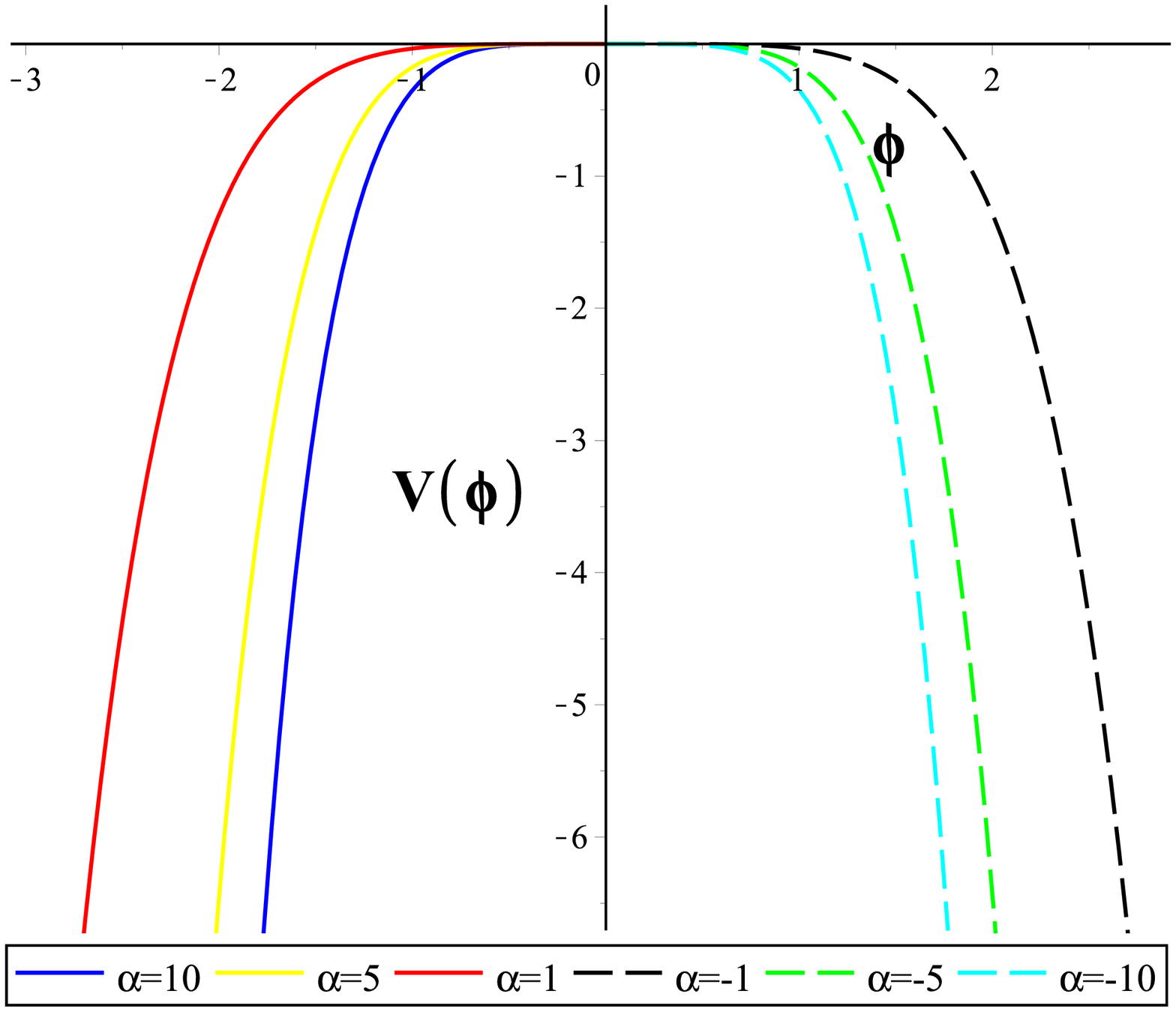}
	\caption{
		The left hand side graphic (a) describe the potential (\ref{pot1}) at $\nu=5$. There are two families: $\phi\in (-\infty,0]$ for which $\alpha>0$ and $\phi\in [0,+\infty)$ for which $\alpha<0$.\\
	The right hand side graphic (b) describe the potential (\ref{pot2}). There are two families: $\phi\in (-\infty,0]$ for which $\alpha>0$ and $\phi\in [0,+\infty)$ for which $\alpha<0$.	  
		\label{Vphi}}
\end{figure}
\newpage
\subsection{Black hole solution}
\label{Sol1}
Considering the action (\ref{action}) with scalar potential (\ref{pot1}) and the metric ansatz
\begin{equation}
ds^{2}=\Omega(x)\left[  -c^{2}f(x)dt^{2}+\frac{\eta^{2}dx^{2}}{f(x)}+d\theta
^{2}+\sin^{2}\theta d\phi^{2}\right]  \label{Ansatz}%
\end{equation}
The equations of motion can be integrated for the conformal factor
\cite{Anabalon:2013sra, Anabalon:2013qua, Acena:2012mr,Acena:2013jya}:
\begin{equation}
\Omega(x)=\frac{\nu^{2}x^{\nu-1}}{\eta^{2}(x^{\nu}-1)^{2}} \label{omega}%
\end{equation}
where the parameters that characterize 
the hairy solutions are $\alpha$ and $\nu$. The solutions for scalar field 
\begin{equation}
\phi(x)=l_{\nu}^{-1}\ln{x}%
\end{equation}
and metric function
\begin{equation}
f(x)=\alpha\biggl{[}\frac{1}{\nu^{2}-4}-\frac{x^{2}}{\nu^{2}%
}\biggl{(}1+\frac{x^{-\nu}}{\nu-2}-\frac{x^{\nu}}{\nu+2}%
\biggr{)}\biggr{]}+\frac{x}{\Omega(x)} \label{f}%
\end{equation}
where $\eta$ is the only integration constant and $l_{\nu}^{-1}=\sqrt{(\nu
	^{2}-1)/2\kappa}$.

The scalar potential given in (\ref{pot1}) and the solution for $\Omega(x), \phi(x), f(x)$ are invariant under the transformation
$\nu\rightarrow-\nu$. The boundary
where $\Omega(x)$ is blowing-up correspond to $x=1$ and the theory
has a standard flat vacuum $V(\phi=0)=0$.
The hairy parameter vary in the range $\nu\in [1,+\infty)$\footnote{Since the potential and the solution is symmetric about $\nu\rightarrow -\nu$, then the behavior is the same if we consider the range $\nu\in (-\infty,-1]$.}, and in the
limit $\nu=1$ one gets $l_{\nu}\rightarrow\infty$ and
$\phi\rightarrow0$ so that the Schwarzschild black hole (asymptotically flat) is smoothly obtained.\\
There are two distinct branches, one that
corresponds to $x\in(0,1]$ and the other one to $x\in[1,\infty)$ --- the curvature singularities are at $x=0$ for the
first branch and $x\rightarrow\infty$ for the second one (these are the
locations where the scalar field is also blowing up).
Considering the change of coordinates\footnote{For the negative branch $x<1$. The procedure for obtaining the coordinate transformation (\ref{trans1}) can be found in \cite{Anabalon:2015xvl}.}
\begin{equation}
x=1-\frac{1}{\eta r}+\frac{(\nu^{2}-1)}{24\eta^{3}r^{3}}\biggl{[}1+\frac
{1}{\eta r}-\frac{9(\nu^{2}-9)}{80\eta^{2}r^{2}}\biggr{]}+O(r^{-6})
\label{trans1}
\end{equation}
we can read off the mass from the sub-leading term of $g_{tt}$:
\begin{equation}
-g_{tt}=f(x)\Omega(x)=1-\frac{\alpha+3\eta^{2}}{3\eta
	^{3}r}+O(r^{-3})\label{Lapse}%
\end{equation}
In the section (\ref{massBh}) we use the quasilocal stress tensor to show that the mass of this 
black hole is
\begin{equation}
m_{1}^{'}=\frac{4\pi c^{2}}{\kappa}\biggl{(}\frac{\alpha+3\eta^{2}}{3\eta^{3}}\biggr{)} 
\end{equation}
where the gravitational radio (or mass parameter) is 
$r_{g}=2 m_{1}^{'}G_{N}/c^{2}$
In the section (\ref{massBh}) we verified that this is the gravitational mass of the black hole
\subsection{Black hole solution $\nu=\infty$}
\label{Sol2}
In \cite{Anabalon:2013qua}, was studied 
the following solution, but here 
we consider the neutral case.\\
One more time, considering the action (\ref{action}) with scalar potential (\ref{pot2}) and the metric ansatz
\begin{equation}
ds^{2}=\Omega(x)\left[  -c^{2}f(x)dt^{2}+\frac{\eta^{2}dx^{2}}{x^{2}f(x)}%
+d\theta^{2}+\sin^{2}{\theta}d\phi^{2}\right]
\label{Ansatz2}%
\end{equation}
The equations of motion can be integrated for the conformal factor
\cite{Anabalon:2013sra, Anabalon:2013qua}:
\begin{equation}
\Omega(x)=\frac{x}{\eta^{2}\left(  x-1\right)  ^{2}} 
\end{equation}
Where $\alpha$ is the parameter that characterizes the
hairy solution. With this choice of the conformal factor is straightforward to obtain the expressions for the scalar field
\begin{equation}
\phi(x)=j^{-1}\ln{x}, \qquad j^{-1}=\frac{1}{\sqrt{2\kappa}}
\end{equation}
and the metric function 
\begin{equation}
f(x)=\alpha\left[  \frac{(x^{2}-1)}{2x}-\ln(x)\right]
+\frac{1}{\Omega(x)}
\label{f1}
\end{equation}
At the boundary $\Omega(x)$ is blowing-up, this correspond to $x=1$. The theory has a standard flat vacuum $V(\phi=0)=0$. There are two distinct branches, one that corresponds to $x\in(0,1]$ and the other one to $x\in[1,\infty)$ --- the curvature singularities are at $x=0$ for the
first branch and $x\rightarrow\infty$ for the second one (these are the
locations where the scalar field is also blowing up). But the difference 
with the before hairy solution
is that does not exist a hairy parameter,
and we can not obtain the Schwarzschild case.\\
Here, $\eta$ is the constant of integration and considering the horizon equation
we can solve it
\begin{equation}
f(\eta,x_{h})=0 \rightarrow \eta^{2}=\frac{x_{h}}{(x_{h}-1)^{2}}\bl{[}\frac{\alpha (2x_{h}\ln{x_{h}}-x_{h}^{2}+1)}{2x_{h}}\br{]}
\label{ecuhori}
\end{equation}
Considering the following asymptotic coordinate transformation\footnote{For the negative branch.}
\begin{equation}
x=1-\frac{1}{\eta r}+\frac{1}{2\eta^{2}r^{2}}-\frac{1}{8\eta^{3}r^{3}}+O(r^{-5})
\label{trans2}
\end{equation}
the gravitational radio $r_{g}$ (or mass parameter)
can be read-off $-g_{tt}$ component in r-coordinate
\begin{equation}
\Omega(x)f(x)=1-\frac{\alpha}{6\eta^{3}r}+O(r^{-3})
\end{equation}
then $r_{g}=\alpha/6\eta^{3}$, and the mass of the hairy black hole is
\begin{equation}
m_{2}^{'}=\frac{4\pi c^{2} }{\kappa}r_{g}=\frac{c^{2}}{2G_{N}}\bl{(}\frac{\alpha}{6\eta^{3}}\br{)}, \qquad 
\end{equation}
In the next section (\ref{massBh}) we verified that this is the gravitational mass of the black hole
\section{Mass of hairy black holes}
\label{massBh}
Similar to holographic formalism
for asymptotic AdS space-times \cite{Myers:1999psa,Anabalon:2015ija,Anabalon:2015xvl,Anabalon:2016izw,Balasubramanian:1999re}, for asymptotically-flat exist an identical proposal given in \cite{Astefanesei:2005ad,Astefanesei:2009wi,Astefanesei:2006zd,Astefanesei:2010bm}. If the 
boundary has the following topology 
$S^{2}\times R\times S^{1}$ the gravitational counterterm is 
\begin{equation}
S^{ct}=-\frac{1}{\kappa}\int_{\pa\mathcal{M}}d^{3}x\sqrt{2\mathcal{R}}~\sqrt{-h}
\end{equation}
The (quasilocal) stress tensor was defined in \cite{Brown:1992br} like
\begin{equation}
\tau_{ab}\equiv\frac{2}{\sqrt{-h}}\frac{\delta S}{\delta h^{ab}}
\end{equation}
where for the total action (that including the boundary terms) given in (\ref{action}) we have \cite{Astefanesei:2005ad}
\begin{equation}
\tau_{ab}=-\frac{1}{\kappa}\left[K_{ab}-h_{ab}K-\Psi(\mathcal{R}_{ab}-\mathcal{R}h_{ab})-h_{ab}\Box\Psi+\Psi_{;ab}\right]
\end{equation}
Considering the foliation ($x=constant$) of the metrics (\ref{Ansatz}) and (\ref{Ansatz2}) (both foliations are similar)
\begin{equation}
h_{ab}dx^{a}dx^{b}=\Omega(x)[-c^{2}f(x)dt^{2}+d\theta^{2}+\sin^{2}{\theta}d\varphi^{2}]
\label{hmetric}
\end{equation}
Where $\mathcal{R}_{ab}$, $\mathcal{R}$ are the Ricci 
tensor and Ricci scalar of the foliation (\ref{hmetric}).
The expression for $\Psi$ was given in \cite{Astefanesei:2006zd}, this is $\Psi=\sqrt{2/\mathcal{R}}$, then
\begin{equation}
\mathcal{R}_{00}=0
\quad,\quad
\mathcal{R}_{\theta\theta}=1
\quad,\quad
\mathcal{R}_{\varphi\varphi}=\sin^2\theta
\quad,\quad
\mathcal{R}=\frac{2}{\Omega(x)}
\quad,\quad
\Psi=\sqrt{\Omega(x)}
\end{equation}
Here $K_{ab}$, $K$ and $n_{a}$ are the
extrinsic curvature, its trace and the normal of time-like hypersurface defined by induced metric (\ref{hmetric}). 
We use the following very useful expressions\footnote{These expressions are correct only when the metric $g_{\mu\nu}$ and the induced metric $h_{ab}$ are diagonal.}
\begin{equation}
K_{ab}=\frac{\sqrt{g^{xx}}}{2}\pa_{x}h_{ab}, \qquad n_{a}=\frac{\delta_{a}^{x}}{\sqrt{g^{xx}}}, \qquad K\sqrt{-h}=n^{a}\pa_{a}\sqrt{-h}
\end{equation}
And is easy to show $T^{\nu}_{tt}$
for solution (\ref{Ansatz}) and $T^{\infty}_{tt}$
for (\ref{Ansatz2})
\begin{equation}
T^{\nu}_{tt}=-\frac{c^{2}}{\kappa}\bl{[}-\frac{(\Omega f)^{'}}{2\eta}\sqrt{\frac{f}{\Omega}}
+\frac{\sqrt{\Omega f}}{2\eta}\bl{(}\frac{3\Omega^{'}f}{\Omega}+f^{'}\br{)}-2\sqrt{\Omega}f\br{]}
\end{equation}
\begin{equation}
T^{\infty}_{tt}=-\frac{c^{2}}{\kappa}\bl{[}-\frac{x(\Omega f)^{'}}{2\eta}\sqrt{\frac{f}{\Omega}}
+\frac{x\sqrt{\Omega f}}{2\eta}\bl{(}\frac{3\Omega^{'}f}{\Omega}+f^{'}\br{)}-2\sqrt{\Omega}f\br{]}
\end{equation}
The conserved quantity associated with the generator of time translations symmetry  $\xi=\pa/c\pa t$ is the energy. For both metrics (\ref{Ansatz}) and (\ref{Ansatz2}) we have
\begin{equation}
E=c^{4}\int d^{2}\Sigma\sqrt{\sigma}m^{a}\xi^{b}T_{ab}=4\pi c^{2} T_{tt}\sqrt{\frac{\Omega}{f}}~\br{\vert}_{x\rightarrow 1} 
\end{equation}
Where $\sigma_{ab}$ is the metric of the transversal section (spherical)
where $\sqrt{\sigma}=\Omega\sin{\theta}$. The normal to foliation $(t=constant)$ is $m_{a}=\delta_{a}^{t}c\sqrt{\Omega f}$. The energy or mass of the hairy black holes (\ref{Ansatz}), (\ref{Ansatz2}) are respectively ($E=m^{'}c^{2}$)
\begin{equation}
m^{'}_{1}=\frac{4\pi c^{2}}{\kappa}\bl{[}\frac{\alpha+3\eta^{2}}{3\eta^{3}}+O(r^{-1})\br{]}_{r=\infty}
\end{equation}
\begin{equation}
m^{'}_{2}=\frac{4\pi c^{2}}{\kappa}
\bl{[}\frac{\alpha}{6\eta^{3}}+O(r^{-1})\br{]}_{r=\infty}
\end{equation}
%
\section{Hamilton-Jacobi method}

In this section, we use the Hamilton-Jacobi method to obtain
the correction to periapsis shift and deflection of light by hairy 
solutions given in (\ref{Sol1})
and (\ref{Sol2}). We considered the intuitive procedure given in the section (\ref{Gr}) 
\subsection{Corrections of hairy solution}
\begin{figure}[h]
	\centering
	\includegraphics[scale=0.39]{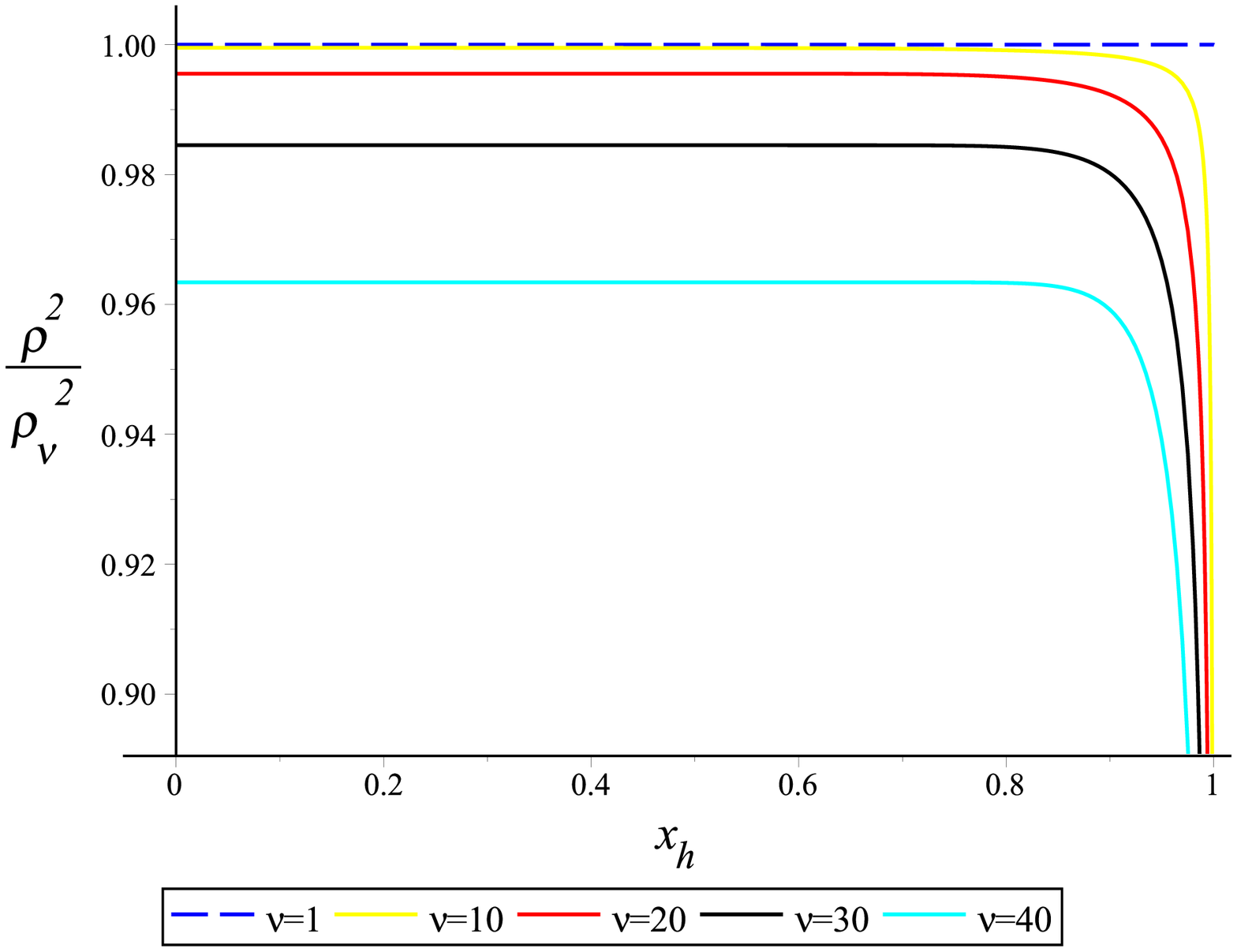}
	\includegraphics[scale=0.39]{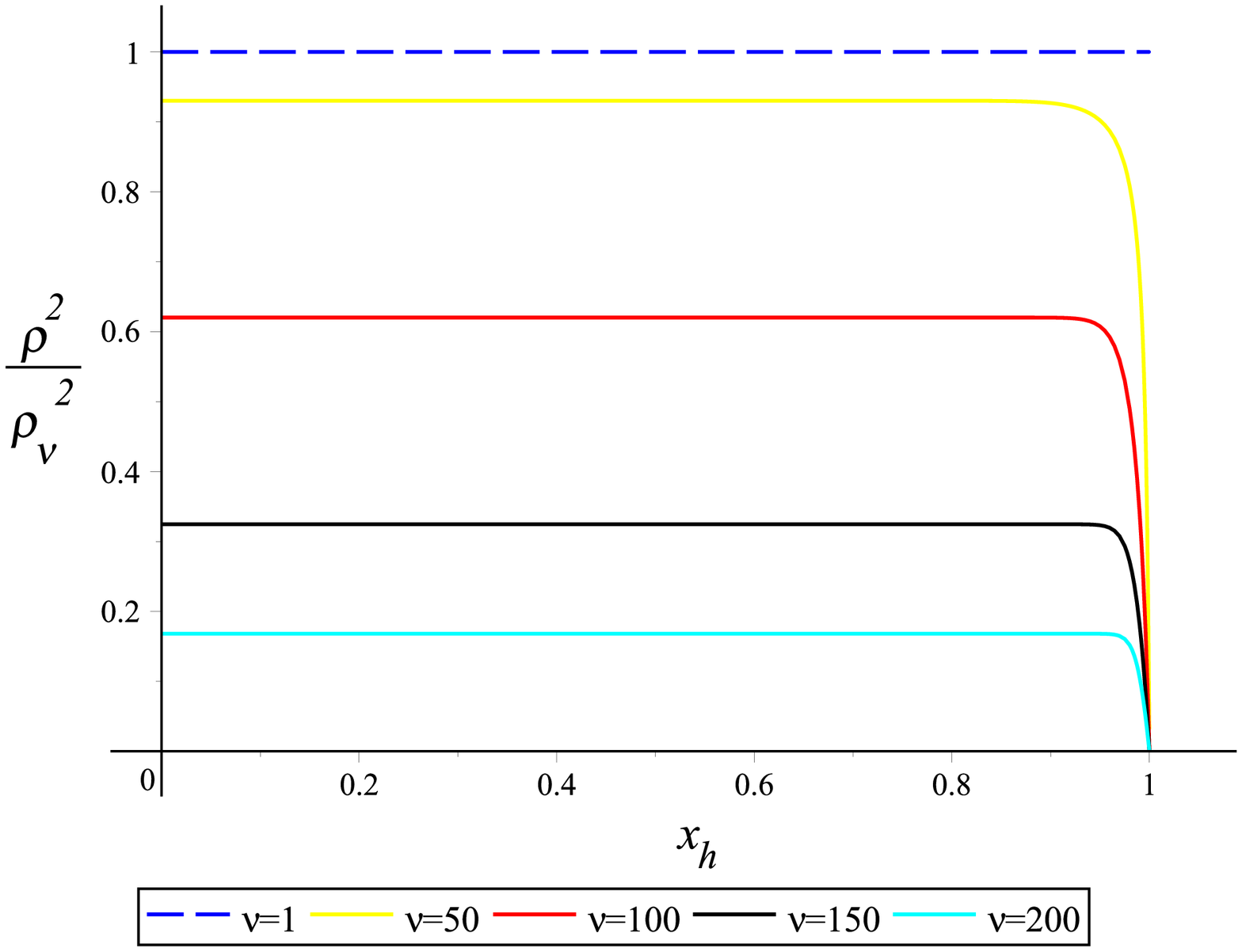}
	\caption{
		The left hand side graphic (e) describe (\ref{rho1}) versus $x_{h}$ for the negative branch, we fix $\varrho=200R_{\odot}$ and $\alpha=10R_{\odot}^{-2}$. And according to our approximation $r_{g}/r<<1$ which means that $x_{h}<<1$, (see the graphic) all these hairy black holes ($\nu>1$) have the same screening factor, which is  $\frac{\varrho^{2}}{\varrho_{\nu}^{2}}<1$.  \\
		The right hand side graphic (f) describe (\ref{rho1}) versus $x_{h}$ for the negative branch, we fix $\varrho=200R_{\odot}$ and $\alpha=10R_{\odot}^{-2}$. And according to our approximation $r_{g}/r<<1$ which means that $x_{h}<<1$, (see the graphic) all these hairy black holes ($\nu>1$) have the same screening factor, which is $\frac{\varrho^{2}}{\varrho_{\nu}^{2}}<1$. But here the screening is larger than the previous case, because the hair is bigger.  
		\label{rhovsX}}
\end{figure}

We study the geodesics and null
geodesic of the metric given in (\ref{Ansatz}). Setting $\theta=\pi /2$,
we have
\begin{equation}
ds^{2}=\Omega(x)\bl{[}-c^{2}f(x)dt^{2}+\frac{\eta^{2}dx^{2}}{f(x)}+d\varphi^{2}\br{]}
\end{equation}
The Hamilton-Jacobi equation for geodesic of some celestial body of mass $m$, that orbit around a hairy black hole is
\begin{equation}
-\frac{1}{c^{2}\Omega f}\bl{(}\frac{\pa I}{\pa t}\br{)}^{2}+\frac{f}{\Omega\eta^{2}}\bl{(}\frac{\pa I}{\pa x}\br{)}^{2}+\frac{1}{\Omega}\bl{(}\frac{\pa I}{\pa\varphi}\br{)}^{2}+m^{2}c^{2}=0
\end{equation}
The ansatz $I=-\mathcal{E}_{0}t+M\varphi+I_{x}(x)$ and its solution is
\begin{equation}
I_{x}(x)=\int\sqrt{\frac{\eta^{2}}{f^{2}}\bl{(}\frac{\mathcal{E}_{0}^{2}}{c^{2}}-m^{2}c^{2}\Omega f\br{)}-\frac{\eta^{2}M^{2}}{f}}~dx
\end{equation}
Considering the coordinate transformation (\ref{trans1}) and similar to Schwarzschild case we need consider $r\rightarrow r+\frac{r_{g}}{2}$
and $\mathcal{E}_{0}=mc^{2}+\mathcal{E}^{'}$, then 
\begin{equation}
x=1-\frac{1}{\eta (r+\frac{r_{g}}{2})}+\frac{(\nu^{2}-1)}{24\eta^{3}(r+\frac{r_{g}}{2})^{3}}\biggl{[}1+\frac
{1}{\eta (r+\frac{r_{g}}{2})}-\frac{9(\nu^{2}-9)}{80\eta^{2}(r+\frac{r_{g}}{2})^{2}}\biggr{]}+O(r^{-6})
\label{transhairy1}
\end{equation}
We can write $I_{x}$ like $I_{r}$ when
$\frac{r_{g}}{r}<<1$
\begin{equation}
I_{r}=\int\sqrt{2m(E-U)+\frac{3\mathcal{A}m^{2}c^{2}r_{g}^{2}}{2r^{2}}-\frac{M^{2}}{r^{2}}+O\bl{(}\frac{r_{g}^{3}}{r^{3}}\br{)}}dr, \qquad \mathcal{A}=1-\frac{\mathcal{E}^{'}(\nu^{2}-1)}{3mc^{2}\eta^{2}r_{g}^{2}}
\label{hairy1}
\end{equation}
Where the expressions for $E$, $U$ was given
in (\ref{EUN}). The new term $\mathcal{A}$ depend on hairy parameter $\nu$ and constant integration $\eta$. We can show that $\mathcal{A}$ give a hairy correction 
to periapsis.  For this purpose, we need expand (\ref{hairy1}) around $r_{g}/r<<1$ (keeping fix $M^{2}/r^{2}$) and $E\approx\mathcal{E}^{'}$, $U\approx U_{N}$
\begin{equation}
\Delta I_{r}=\int_{r_{min}}^{r_{max}}\sqrt{2m(\mathcal{E}^{'}-U_{N})-\frac{M^{2}}{r^{2}}}+\frac{3m^{2}c^{2}r_{g}^{2}\mathcal{A}}{4}\int_{r_{min}}^{r_{max}}\frac{1}{r^{2}}\bl{[}\sqrt{2m(\mathcal{E}^{'}-U_{N})-\frac{M^{2}}{r^{2}}}\br{]}^{-1}dr+O\bl{(}\frac{r_{g}^{4}}{r^{4}}\br{)}
\end{equation}
\begin{equation}
\Delta I_{r}=\Delta I_{r}^{(0)}-\frac{3m^{2}c^{2}r_{g}^{2}\mathcal{A}}{4M}\frac{\pa\Delta I_{r}^{(0)}}{\pa M}+O\bl{(}\frac{r_{g}^{4}}{r^{4}}\br{)}
\end{equation}
And is easily to show, 
\begin{equation}
\Delta\varphi\approx 2\pi+\frac{3\pi m^{2}c^{2}r_{g}^{2}\mathcal{A}}{2M^{2}}, \qquad \mathcal{A}=1+\frac{(\nu^{2}-1)}{4a}\bl{(}\frac{\eta}{\alpha+3\eta^2}\br{)}
\label{varphi1}
\end{equation}
when $\nu=1$, we have $\mathcal{A}=1$,  e.i. the Schwarzschild case given in (\ref{Schwaphi}).
Where the orbital parameter $a$ is the semi-major axis.\footnote{According to astronomical observations \cite{Eisenhauer:2005cv,Ghez:2003qj}, nine stars are currently known that orbit around of Sagittarius $A^{*}$, including $S1$, $S2$, $S8$. And its semi-major axis oscillate between $a\sim 900-3500~AU$. The astronomical unit $(AU)$ or radius of Earth's orbit, and it is equivalent to $1.495\times 10^{11}~m$.\\
According to \cite{Ghez:1998ph,Ghez:2003qj,Schodel:2002vg} we can show $r_{g}=\frac{2m^{*}G_{N}}{c^{2}}\approx 0.1AU$, where $m^{*}$ is the mass of Sagittarius $A^{*}$, then $r_{g}<<a$.} 
If we consider the horizon
equation $f(x_{h},\eta,\alpha)=0$ and solve $\eta=\eta (x_{h},\alpha)$, we can
write the hairy correction $\mathcal{A}$ as a function $\mathcal{A}=\mathcal{A}(a,\nu,x_{h},\alpha)$
\begin{figure}[h]
	\centering
	\includegraphics[scale=0.39]{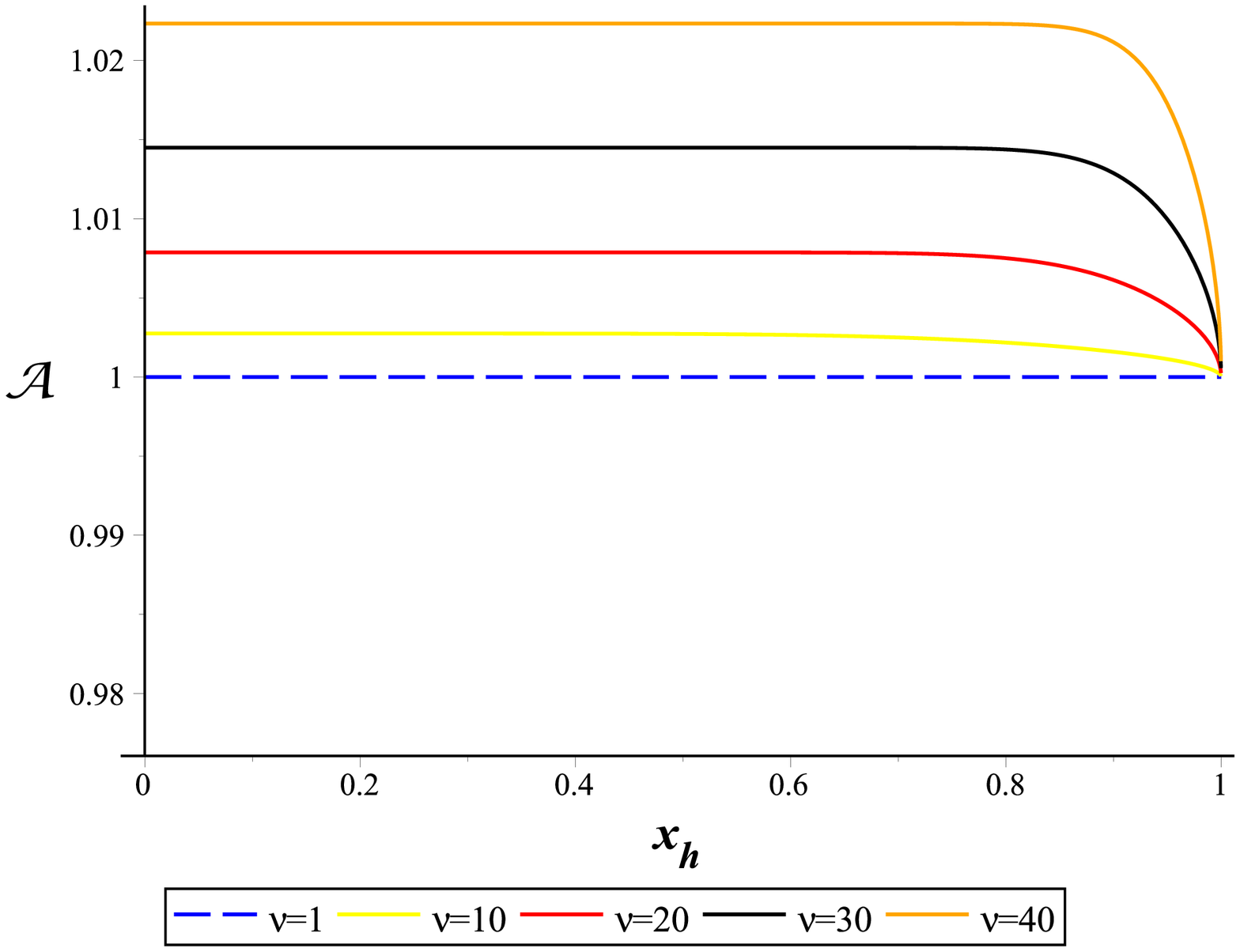}
	\includegraphics[scale=0.39]{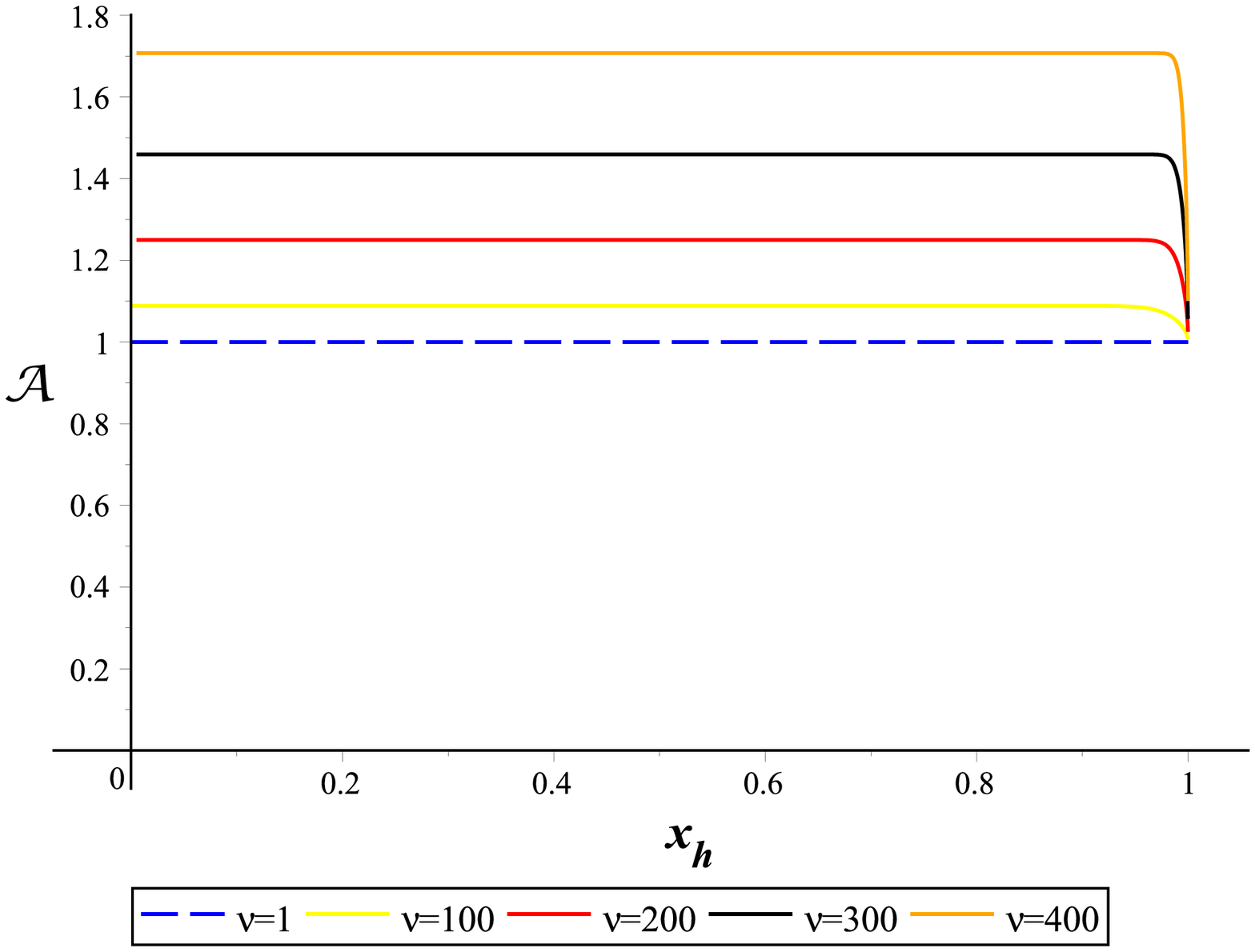}
	\caption{
		The left hand side graphic (c) describe (\ref{varphi1}) versus $x_{h}$ for the negative branch, we fix $a=2000~AU$ and $\alpha=2~AU^{-2}$. And according to our approximation $r_{g}/r<<1$ which means that $x_{h}<<1$. That is, in the graphic all these hairy black holes ($\nu>1$) have the same corrective factor, which is  $\mathcal{A}>1$.  \\
		The right hand side graphic (d) describe (\ref{varphi1}) versus $x_{h}$ for the negative branch, we fix $a=2000~AU$ and $\alpha=2~AU^{-2}$. And according to our approximation $r_{g}/r<<1$ which means that $x_{h}<<1$. That is, in the graphic all these hairy black holes ($\nu>1$) have the same corrective factor, which is  $\mathcal{A}>1$. But here the correction is larger than the previous case, because the hair is bigger.  
		\label{AvsX}}
\end{figure}
\newpage
Similar to Schwarzschild case, we can construct the eikonal equation, whose solution is
\begin{equation}
\psi=-\omega_{0}t+M\varphi+\psi_{x}
\end{equation}
\begin{equation}
\psi_{x}=\frac{\omega_{0}}{c}\int{\sqrt{1-\varrho^{2}f}}\frac{\eta}{f}dx, \qquad \varrho=\frac{Mc}{\omega_{0}}
\end{equation}
The equation of light trajectory is given by\footnote{Is easy to show that the Schwarzschild case ($\nu=1$) given in (\ref{lightrayec}) can be obtained when
	\begin{equation}
	x=1-\frac{1}{\eta r}, \qquad \Omega f=1-\frac{r_{g}}{r}, \qquad \Omega=r^{2}, \qquad \eta dx=\frac{dr}{r^{2}}
	\end{equation}}
\begin{equation}
\varphi(x)=\int\frac{\eta dx}{\sqrt{\frac{1}{\varrho^{2}}-f}}+cte
\end{equation}
We need to consider the hairy-gravity corrections, considering one more time
the coordinate transformations given in (\ref{transhairy1}) we can show
\begin{equation}
\psi_{r}\approx\frac{\omega_{0}}{c}\int\sqrt{1+\frac{2r_{g}}{r}-\frac{\varrho_{\nu}^{2}}{r^{2}}}dr
\end{equation}
where the impact parameter $\varrho$ has a new hairy correction 
\begin{equation}
\varrho^{2}_{\nu}=\varrho^{2}+\frac{\nu^{2}-1}{4\eta^{2}}
\end{equation}
Now, expanding around $r_{g}/r<<1$, keeping fix $\varrho_{\nu}^{2}/r^{2}$
\begin{equation}
\psi_{r}\approx\frac{\omega_{0}}{c}\int\sqrt{1-\frac{\varrho^{2}}{r^{2}}}dr+\frac{\omega_{0}r_{g}}{c}\int\frac{dr}{\sqrt{r^{2}-\varrho_{\nu}^{2}}}
\end{equation}
In the first term of the expansion of $\psi_{r}$ we have $\sqrt{1-\frac{\varrho^{2}}{r^{2}}-\frac{\nu^{2}-1}{4\eta^{2}r^{2}}}\approx\sqrt{1-\frac{\varrho^{2}}{r^{2}}}$, this is because
if we consider the horizon equation we have $\eta=\eta (x_{h},\alpha,\nu)$, and you can prove that for $x_{h}<<1$ ($r_{g}<<r$) we have $\eta^{2}$ large, then $\frac{3}{4r^{2}\eta^2}\approx 0$. But for
the second term we have only $\sqrt{r^{2}-\varrho^{2}-\frac{\nu^{2}-1}{4\eta^{2}}}$, from which $\frac{\nu^{2}-1}{4\eta^{2}}$ is relevant. This is precisely the relevant hairy correction
\begin{equation}
\Delta\varphi=-\frac{\pa\Delta\psi_{r}^{(0)}}{\pa M}+\frac{2r_{g}Mc}{\omega_{0}\varrho_{\nu}^{2}}\frac{1}{\sqrt{1-\frac{\varrho_{\nu}^{2}}{R^{2}}}}
\end{equation}
when $R\rightarrow\infty$ 
\begin{equation}
\Delta\varphi\approx\pi+\frac{2r_{g}}{\varrho}~\frac{\varrho^{2}}{\varrho_{\nu}^{2}}
\end{equation}
And one more time when $\nu=1$
we obtain the Schwarzschild\footnote{For a ray of light passing near the edge of the sun we obtain $1.75^{''}$.} case giving in (\ref{Schwalighdeflex}). Using the horizon equation $f(\eta,x_{h},\nu,\alpha)=0$ we have $\eta=\eta (x_{h},\nu,\alpha)$
\begin{equation}
\frac{\varrho_{\nu}^{2}}{\varrho^{2}}\br{\vert}_{(x_{h},\nu,\alpha)}=1+\frac{(\nu^{2}-1)}{4\eta^{2}\varrho^{2}}
\label{rho1}
\end{equation}
To graph $\varrho^{2}/\varrho_{\nu}^{2}$ vs $x_{h}$, we consider $\varrho$ like multiples of $R_{\odot}$ (sun radius). For example 
the stars early-B hyper-giants (BHGs),
has radios in the range of  $\varrho=63.5R_{\odot}-246R_{\odot}$ \cite{Clark:2012ne}. 
The figures in (\ref{rhovsX}) show that the scalar field (hair) screening the effect of gravitational field, causing the deviation of light to be smaller
respect to Schwarzschild case ($\nu=1$)
\newpage
\subsection{Corrections of hairy solution $\nu=\infty$} 
We study the geodesics and null
geodesics of the metric given in (\ref{Ansatz2}). Setting $\theta=\pi /2$, we have
\begin{equation}
ds^{2}=\Omega(x)\bl{[}-c^{2}f(x)dt^{2}+\frac{\eta^{2}dx^{2}}{x^{2}f(x)}+d\varphi^{2}\br{]}
\end{equation}
The Hamilton-Jacobi equation is
\begin{equation}
-\frac{1}{c^{2}\Omega f}\bl{(}\frac{\pa I}{\pa t}\br{)}^{2}+\frac{x^{2}f}{\Omega\eta^{2}}\bl{(}\frac{\pa I}{\pa x}\br{)}^{2}+\frac{1}{\Omega}\bl{(}\frac{\pa I}{\pa\varphi}\br{)}^{2}+m^{2}c^{2}=0
\end{equation}
the ansatz $I=-\mathcal{E}_{0}t+M\varphi+I_{x}(x)$,
and the solution for $I_{x}(x)$ is
\begin{equation}
I_{x}=\int\sqrt{\frac{\eta^{2}}{x^{2}f^{2}}\bl{(}\frac{\mathcal{E}_{0}}{c^{2}}-m^{2}c^{2}\Omega f\br{)}-\frac{\eta^{2}M^{2}}{x^{2}f}}~dx
\end{equation}
Considering the coordinate transformation (\ref{trans2}), and in a similar form to Schwarzschild case, we need to consider $r\rightarrow r+\frac{r_{g}}{2}$
and $\mathcal{E}_{0}=mc^{2}+\mathcal{E}^{'}$, then, we can write $I_{x}$ like $I_{r}$ when
$\frac{r_{g}}{r}<<1$
\begin{equation}
I_{r}=\int\sqrt{2m(E-U)+\frac{3\mathcal{A}m^{2}c^{2}r_{g}^{2}}{2r^{2}}-\frac{M^{2}}{r^{2}}+O\bl{(}\frac{r_{g}^{3}}{r^{3}}\br{)}}~dr, \qquad \mathcal{A}=1-\frac{\mathcal{E}^{'}}{mc^{2}\eta^{2}r_{g}^{2}}
\label{hairy2}
\end{equation}
Where the expressions for $E$, $U$ was given
in (\ref{EUN}). The new term $\mathcal{A}$ depend on constant integration $\eta$. We show that $\mathcal{A}$ give a hairy correction 
to periapsis shift. For these purpose we need expand (\ref{hairy2}) around $r_{g}/r<<1$ (keeping fix $M^{2}/r^{2}$) and $E\approx\mathcal{E}^{'}$, $U\approx U_{N}$.
And we have,
\begin{equation}
\Delta\varphi\approx 2\pi+\frac{3\pi m^{2}c^{2}r_{g}^{2}\mathcal{A}}{2M^{2}}
\label{varphi2}
\end{equation}
where the orbital parameter $a$ is the semi-major axis\footnote{Here we use the horizon equation given in (\ref{ecuhori}).},
\begin{equation}
\mathcal{A}(a,\alpha,x_{h})=1+\frac{3\sqrt{\alpha}}{2a}~\sqrt{\frac{2x_{h}\ln{x_{h}}-x_{h}^{2}+1}{2(x_{h}-1)^{2}}}
\label{A2}
\end{equation}
In this case, there are not a hairy parameter such that $\mathcal{A}=1$,
the unique form to obtain $\mathcal{A}=1$ is when $\alpha=0$, but according to
(\ref{f}) and (\ref{f1}), these case correspond to a naked singularity.
Even if we consider black holes asymptotically AdS~\footnote{The hairy black holes (asymptotically AdS) constructed in \cite{Anabalon:2012dw,Anabalon:2013qua,Acena:2012mr,Acena:2013jya}, has more general  scalar potentials which consist of two parts $V(\phi)\sim \alpha (\ldots)+\Lambda (\ldots)$, where $\Lambda=-3/l^{2}$ is the cosmological constant, and when $\alpha=0$ has a new interesting interpretations, like domain wall. Where its structure is ensured for the existence of potential $V(\phi)\sim \Lambda (\ldots)$ \cite{Anabalon:2013eaa}.}, the special case $\alpha=0$ is a naked singularity. Is clear that $\alpha$ has an important role to existence of regular horizon. The condition $\alpha\neq 0$ ensure the existence of scalar potentials  given in (\ref{pot1}), (\ref{pot2}), and this means that if $\alpha=0$, there are not self-interaction of the scalar field that ensures its stability and at the same time, we have a naked singularity.
\begin{figure}[h]
	\centering
	\includegraphics[scale=0.39]{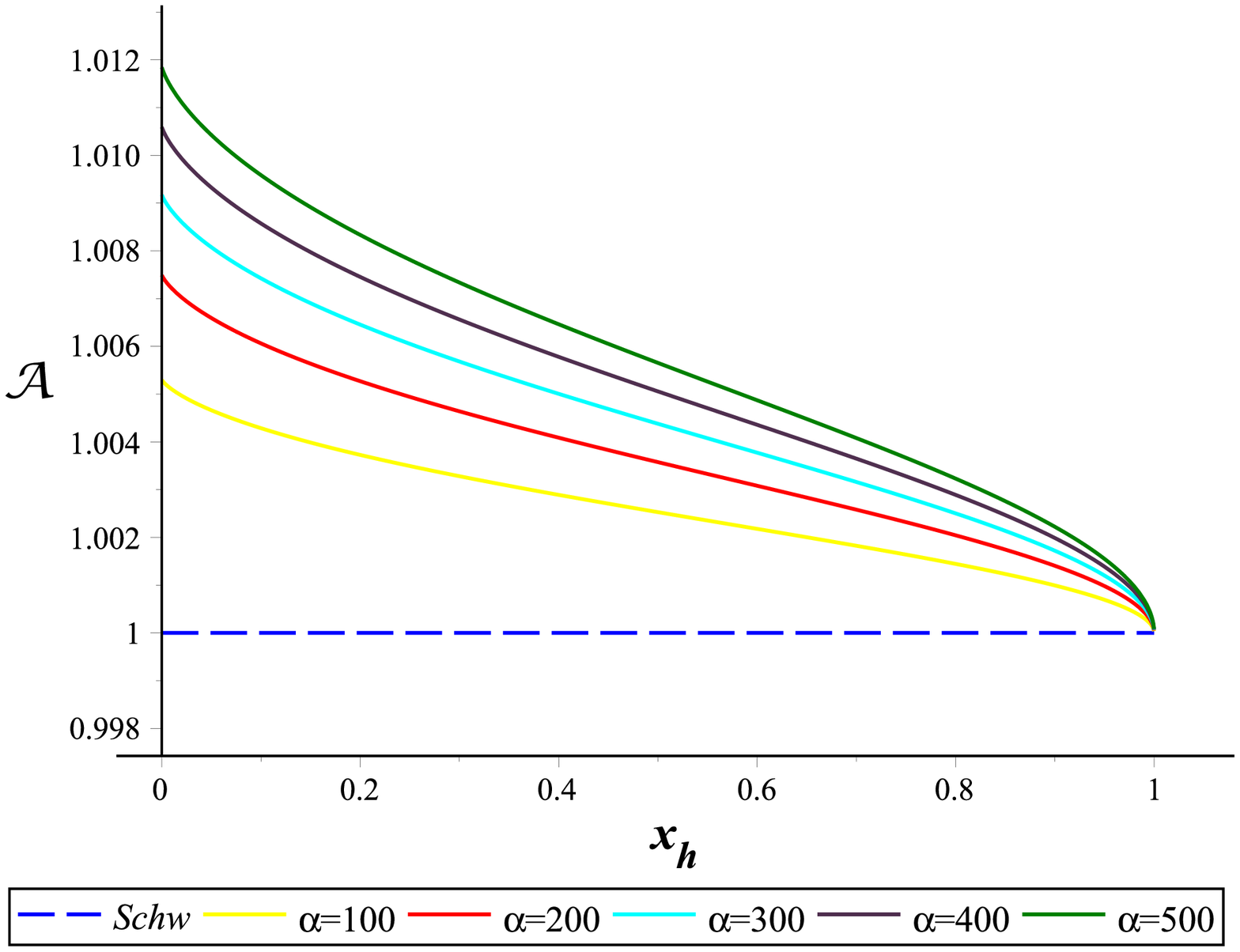}
	\includegraphics[scale=0.39]{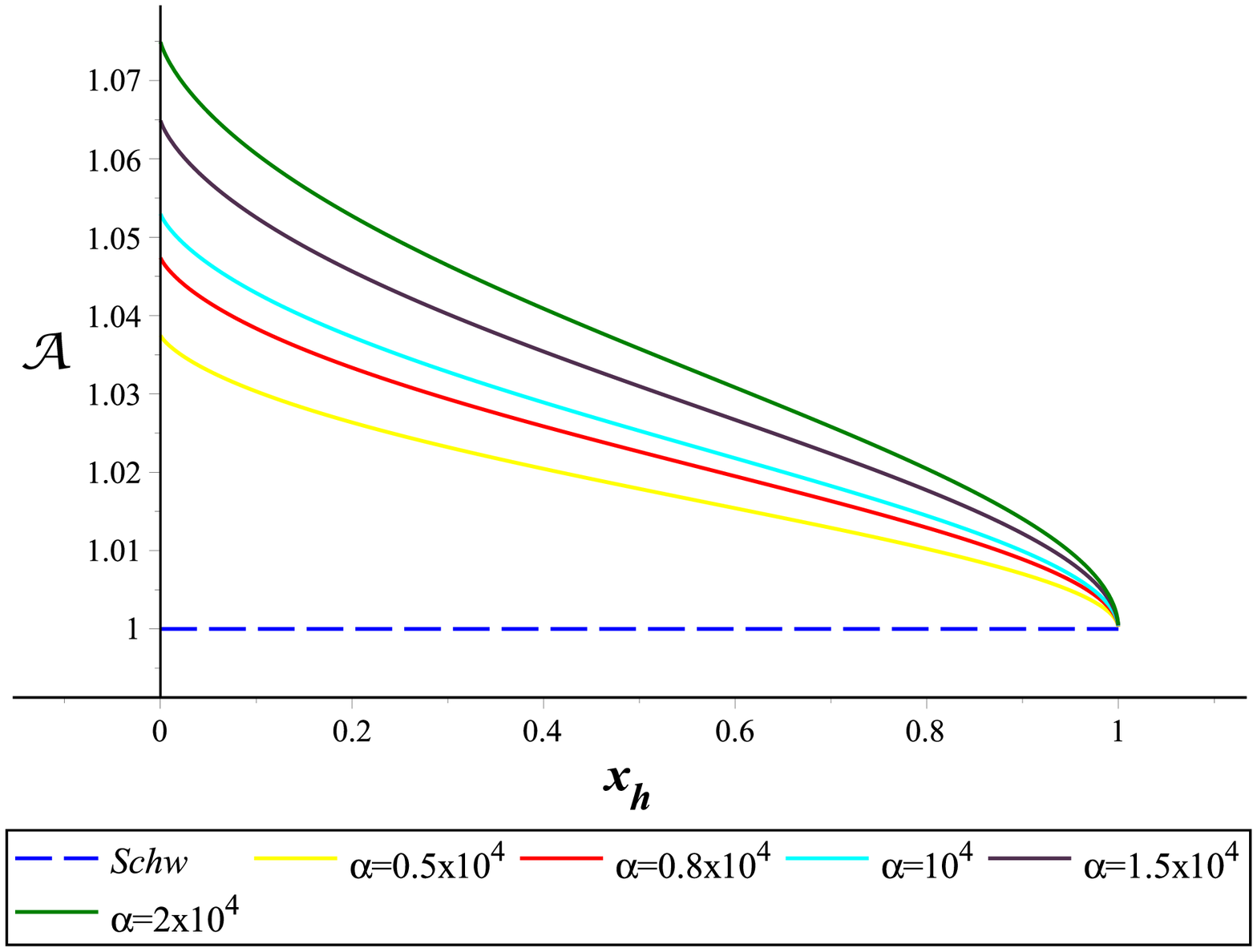}
	\caption{
		The left hand side graphic (g) describe (\ref{A2}) versus $x_{h}$ for the negative branch, we fix $a=2000~AU$ and $\alpha=(100-500)~AU^{-2}$. We put the case $Schw$ (Schwarzschild) to compare. And according to our approximation $r_{g}/r<<1$ which means that $x_{h}<<1$. That is, in the graphic all these hairy black holes have different (small) corrective factors, which is approximate $\mathcal{A}\sim 1$.  \\
		The right hand side graphic (h) describe (\ref{A2}) versus $x_{h}$ for the negative branch, we fix $a=2000~AU$ and $\alpha=(0.5-2)\times10^{4}~AU^{-2}$. We put the case $Schw$ (Schwarzschild) to compare. And according to our approximation $r_{g}/r<<1$ which means that $x_{h}<<1$. That is, in the graphic, all these hairy black holes have different corrective factors. But here the correction is larger than the previous case because $\alpha$ is bigger.  
		\label{AvsX1}}
\end{figure}
\newpage
To describe the light trajectory 
we use the eikonal equation. The ansatz 
is $\psi=-\omega_{0}t+M\varphi+\psi_{x}$, and
the solution for $\psi_{x}$ is
\begin{equation}
\psi_{x}=\frac{\omega_{0}}{c}\int{\sqrt{1-\varrho^{2}f}}\frac{\eta}{xf}dx, \qquad \varrho=\frac{Mc}{\omega_{0}}
\end{equation}
The equation of light trajectory is given by
\begin{equation}
\varphi(x)=\int\frac{\eta dx}{x\sqrt{1/\varrho^{2}-f}}+cte
\end{equation}
But in here, we can not obtain the Schwarzschild as a smooth limit.
To get the hairy-gravity corrections, we need to consider one more time
the coordinate transformations given in (\ref{trans2}) we can show, for $\frac{r_{g}}{r}<<1$  
\begin{equation}
\psi_{r}\approx\frac{\omega_{0}}{c}\int\sqrt{1+\frac{2r_{g}}{r}-\frac{\varrho_{\alpha}^{2}}{r^{2}}}dr
\end{equation}
where the impact parameter $\varrho$ has a new hairy term 
\begin{equation}
\varrho^{2}_{\alpha}=\varrho^{2}+\frac{3}{4\eta^{2}}
\end{equation}
Now, expanding around $r_{g}/r<<1$, keeping fix $\varrho_{\nu}^{2}/r^{2}$, we have
\begin{equation}
\psi_{r}\approx\frac{\omega_{0}}{c}\int\sqrt{1-\frac{\varrho^{2}}{r^{2}}}dr+\frac{\omega_{0}r_{g}}{c}\int\frac{dr}{\sqrt{r^{2}-\varrho_{\alpha}^{2}}}
\end{equation}
In the first term of the expansion of $\psi_{r}$ we have $\sqrt{1-\frac{\varrho^{2}}{r^{2}}-\frac{3}{4\eta^{2}r^{2}}}\approx\sqrt{1-\frac{\varrho^{2}}{r^{2}}}$, this is because
if we consider $\eta=\eta (x_{h},\alpha)$ given in (\ref{ecuhori}), you can prove that for $x_{h}<<1$ ($r_{g}<<r$) we have $\eta^{2}$ large, then $\frac{3}{4r^{2}\eta^2}\approx 0$. But for
the second term, we have only $\sqrt{r^{2}-\varrho^{2}-\frac{3}{4\eta^{2}}}$, from which $\frac{3}{4\eta^{2}}$ is not small enough. This is precisely the relevant hairy correction. Working in the same way as in the previous section, we can show that 
\begin{equation}
\Delta\varphi=\pi+\frac{2r_{g}}{\varrho}~\frac{\varrho^{2}}{\varrho_{\alpha}^{2}}; \qquad \frac{\varrho_{\alpha}^{2}}{\varrho^{2}}\br{\vert}_{(x_{h},\alpha)}=1+\frac{3}{2\alpha\varrho^{2}}~\frac{(x_{h}-1)^{2}}{2x_{h}
	\ln{x_{h}-x_{h}^{2}+1}}
\label{rho2}
\end{equation}
\begin{figure}[h]
	\centering
	\includegraphics[scale=0.39]{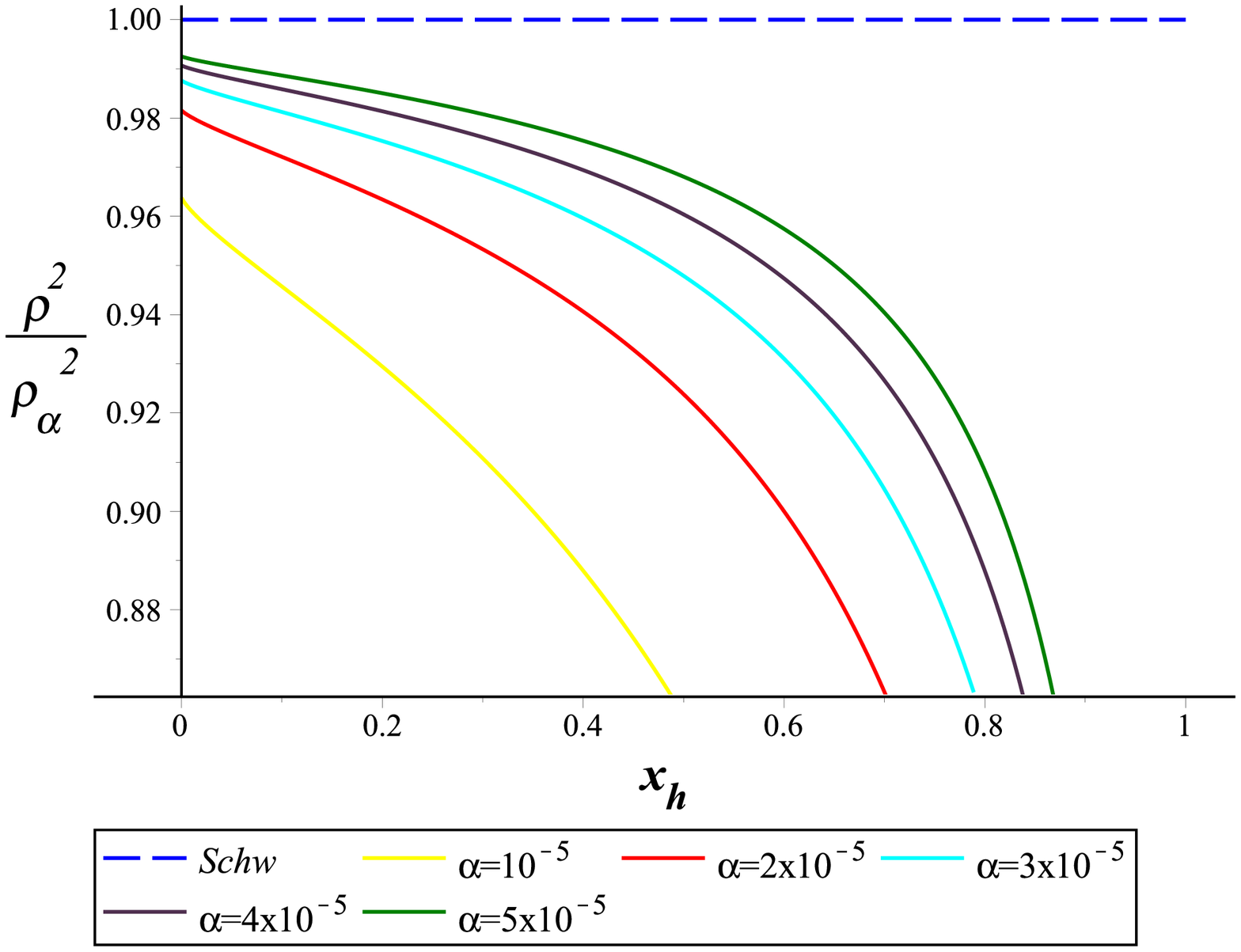}
	\includegraphics[scale=0.39]{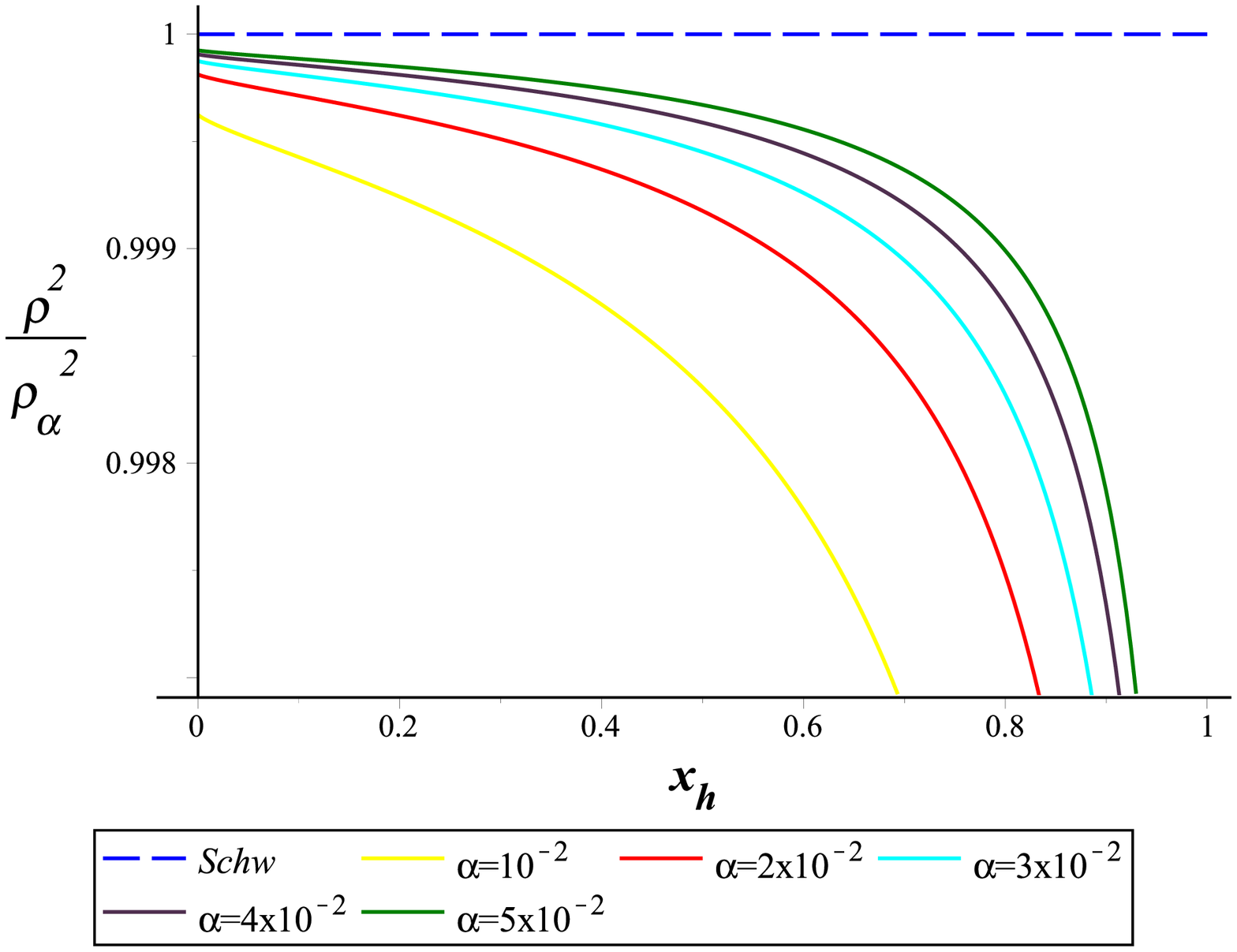}
	\caption{
		The left hand side graphic (i) describe (\ref{rho2}) versus $x_{h}$ for the negative branch, we fix the impact parameter like $\varrho=2000R_{\odot}$. And according to our approximation $r_{g}/r<<1$ which means that $x_{h}<<1$, (see the graphic) these hairy black holes has different screening factors $\frac{\varrho^{2}}{\varrho_{\nu}^{2}}<1$, such that $\alpha\approx 0$.  \\
		The right-hand side graphic (k) describe (\ref{rho2}) versus $x_{h}$ for the negative branch, we fix $\varrho=2000R_{\odot}$. And according to our approximation, $r_{g}/r<<1$ which means that $x_{h}<<1$, (see the graphic) these hairy black holes has small screening factor. Comparing with the before case, the screening is small, then $\frac{\varrho^{2}}{\varrho_{\nu}^{2}}\sim \frac{1}{\alpha}$ at fixed $\varrho$.   
		\label{rhovsX2}}
\end{figure}

\newpage
\section{Conclusions}
  
Sagittarius $A^{*}$ is a stellar system
located in the center of our galaxy, it consists of a supermassive black hole, where until now the orbits of nine stars were studied \cite{Eisenhauer:2005cv,Ghez:2003qj,Ghez:1998ph,Schodel:2002vg,Clark:2012ne}. This black hole has a disk accretion, and there are indications that dark matter influences the gravitational dynamics of these stars \cite{Ghez:1998ph}. In 2004, 
source IRS 13 black hole was discovered 
(and other stars that orbit it) that orbit
around of Sagittarius $A^{*}$. The principal idea of the present paper is that
these hairy solutions (\ref{Sol1}) and (\ref{Sol2}) could be an effective model 
to describe a more complicated gravitational system. Our idea is to consider this gravitational system (for example Sagittarius $A^{*}$) as a black hole with scalar hair,
specially the solution (\ref{Sol1}) has
a hairy parameter $\nu$, it can be adjusted in such a way as to describe for example the Sagittarius $A^{*}$. We propose (like an elemental model) to 
use the results for periapsis
shift given in (\ref{varphi1}),
(\ref{varphi2}),
and made astronomical observations to the stars that orbit Sagittarius $A^{*}$~\footnote{There is no concrete evidence of the rotation Sagittarius $A^{*}$ but in general, all observed the black holes rotate. We consider the hairy black hole as a simple static model of the stationary system of Sagittarius $A^{*}$, assuming that its rotation is slow enough.}, estimate its shift periapsis and thereby set the parameter of hair $\nu$ (and $\alpha$). In a similar form maybe is possible use the equation for deflection of light given in (\ref{rho1}), (\ref{rho2}). 
An interesting future direction is, 
study and classify the different orbits of these hairy black holes.  

\section{Acknowledgments}
D. Choque would like to thank Dumitru Astefanesei for interesting discussions. D. Choque acknowledge the hospitality of the Universidad Nacional de San Antonio Abad del Cusco (UNSAAC) during the stages of this research. This work has been done with support from the 047-2017-FONDECYT-DE. The UNSAAC is funded by the Peruvian Government through Financing Program of CONCYTEC.

\end{document}